\documentclass[final,3p,times]{elsarticle}

 \usepackage{graphics,epsfig,subfigure}

\usepackage{amssymb}
\usepackage{cases}

\usepackage{amstext}
\usepackage{color}
\usepackage{array}
\usepackage{multirow}
\usepackage{hyperref}
\newcommand\beq{\begin{equation}}
\newcommand\eeq{\end{equation}}
\newcommand\beqn{\begin{eqnarray}}
\newcommand\eeqn{\end{eqnarray}}
\newcommand\nn{\nonumber}
\newcommand\fc{\frac}
\newcommand\lt{\left}
\newcommand\rt{\right}
\newcommand\pt{\partial}
 \biboptions{comma,square,compress,numbers,sort}

\journal{Physics Letters B}

\begin{document}

\begin{frontmatter}



\title{Topological approach to derive the global Hawking temperature of (massive) BTZ black hole}


\author[label1]{Yu-Peng Zhang}
 \author[label1]{Shao-Wen Wei}
  \author[label1,label2]{Yu-Xiao Liu\corref{cor1}}
  \ead{liuyx@lzu.edu.cn}
  \cortext[cor1]{The corresponding author.}

\address[label1]{Institute of Theoretical Physics \& Research Center of Gravitation, Lanzhou University, Lanzhou 730000, China\\
Joint Research Center for Physics, Lanzhou University and Qinghai Normal University, Lanzhou 730000 and Xining 810000, China}
\address[label2]{Key Laboratory for Magnetism and Magnetic of the Ministry of Education, Lanzhou University, Lanzhou 730000, China}

\begin{abstract}
In this paper, we study the Hawking temperature of the BTZ black hole based on the purely topological method proposed by Robson, Villari, and Biancalana (RVB) [Phys. Rev. D 99, 044042 (2019)]. The Hawking temperature of the charged rotating BTZ black hole can be accurately derived by this topological method. We also calculate the Hawking temperature of the BTZ black hole in massive gravity. Because the metric function of the BTZ black hole in massive gravity has a mass term, the corresponding Hawking temperature cannot be derived unless an integral constant is added.

\end{abstract}

\begin{keyword}
Topology \sep Euler characteristic \sep Hawking temperature

\end{keyword}

\end{frontmatter}



\section{Introduction}

It is well known that the gravitational waves from merging binary black holes and inspiraling neutron star binaries have been directly detected by the Laser Interferometer Gravitational-Wave Observatory (LIGO) scientific Collaboration and Virgo Collaboration \cite{Abbott2016a,Abbott2016,Abbott2017,Abbott:2017oio,Abbott:2017b}, which provides a powerful evidence for the existence of black holes. In classical viewpoint, a black hole is always considered as an extreme object and nothing can escape from it. While Hawking and Bekenstein found that a black hole can possess temperature $T$ and entropy $S$ \cite{Bekenstein1973a,Bekenstein1973b,Hawking1975}, and a black hole system can be treated as a thermodynamical system \cite{Gibbons1977a,Unruh1976,Damour1976,Gibbons1977b}. The strong gravitational system can be mapped to a thermodynamic system, and the corresponding relationship was proposed by Jacobson \cite{Jacobson1995}.

The topological properties of a black hole can be characterized by the topological invariant Euler characteristic $\chi$ \cite{Gibbons1995,Gibbons1979,Eguchi1980,Liberati1997}. Recently, Robson, Villari, and Biancalana \cite{Robson2018} showed that the Hawking temperature of a black hole is closely related to its topology, and they proposed a topological method related to the Euler characteristic $\chi$ for obtaining the Hawking temperature. This method has been successfully applied to four-dimensional Schwarzschild-de Sitter black holes \cite{Robson2019}, anti-de Sitter black Holes \cite{Robson2019b}, and other Schwarzschild-like or charged black holes \cite{Saka2019}.

On the other hand, it is interesting to check whether the RVB method is applicable to lower-dimensional black holes. One particular case of great interest is the three-dimensional Ba$\tilde{\text{n}}$ados-Teitelboim-Zanelli (BTZ) black hole \cite{Banados1992}, which is very useful to understand gravitational interactions in low-dimensional spacetime \cite{Witten2007}.

After a detailed calculation, we find that the Hawking temperature of the charged and spinning BTZ black hole in general relativity is derived perfectly by using the RVB method. However, for the BTZ black hole in massive gravity, the corresponding Hawking temperature can not be obtained unless an integral constant is added, which is mainly resulted from the massive term. Moreover, starting from an effective two-dimensional black hole metric, we clearly show that the Hawking temperature can be exactly reproduced by using the RVB method.

This paper is organized as follows. In Section \ref{Deriv}, we give a brief review of the RVB method. In Section \ref{btz1}, we use the RVB method to obtain the temperature of the charged and rotating BTZ black hole cases in general relativity. For the BTZ black hole in massive gravity, its Hawking temperature is also reproduced in Section \ref{btz2}. Finally, a brief summary and conclusion will be given in Sec. \ref{Conclusion}.

\section{RVB method} \label{Deriv}

Very recently, Robson, Villari, and Biancalana proposed a powerful formula to investigate the Hawking temperature of a black hole. They showed that the Hawking temperature of a black hole can be topologically derived. This method was also found to be effective for any coordinate system. A black hole is a very special system and has a topological invariant, the Euler characteristic $\chi$ \cite{Gibbons1995,Gibbons1979,Eguchi1980,Liberati1997}. There are many black hole systems described by various kinds of complicated metrics. We usually need highly nontrivial coordinate transformations to study them. But now the thing is different, the Hawking temperature of a black hole can be obtained by using the topological formula related to the Euler characteristic $\chi$.

In the RVB method, one uses the Euclidean geometry of the 2-dimensional spacetime to derive Hawking temperature without losing any information of the higher-dimensional space. The Hawking temperature of a two-dimensional black hole can be derived by the RVB method \cite{Robson2018,Robson2019}
\begin{equation}
T_{\text{H}}=\frac{\hbar c}{4\pi \chi k_{\text{B}}}\Sigma_{j\leq\chi}\int_{r_{h_j}}{\sqrt{|g|}\mathcal{R}dr}.
\label{temperature}
\end{equation}
Here the parameters $\hbar$, $c$, and $k_{\text{B}}$ are the Planck constant, speed of light, and Boltzmann's constant, respectively. And $g$ is the metric determinant,  $r_{h_j}$ is the $j$-th killing horizon.  In this paper, we set the parameters $\hbar=1$, $c=1$, and $k_{\text{B}}=1$. The function $\mathcal{R}$ is the Ricci scalar of the two-dimensional spacetime. The parameter $\chi$ stands for the Euler characteristic of the Euclidean geometry and relates to the number of the Killing horizons. $\Sigma_{j\leq\chi}$ denotes the summation over the Killing horizons.

The Euler characteristic $\chi$ of a black hole is a topological invariant related to the structure of the manifold. Here, the spacetime of a black hole with an outer boundary can be identified as compact manifold. The Euler characteristic $\chi$ in a closed $n$ (even)-dimensional manifold $M^n$ can be defined as \cite{Morgan1993}
\begin{equation}
\chi=\frac{2}{\text{area}(S^n)}\int_{M^n}\sqrt{|g|} d^nx  G,
\label{origin1}
\end{equation}
where $\text{area}(S^n)$ is the surface area of the unit $n$-sphere, $G$ is a density and defined by using the Riemann tensor of the manifold
\begin{eqnarray}
G&=&\frac{1}{2^{n/2}n!g}\epsilon^{i_1\cdots i_n} \epsilon^{j_1\cdots j_n}
     \mathcal{R}_{i_1 i_2 j_1 j_2}\mathcal{R}_{i_3 i_4 j_3 j_4}\cdots \mathcal{R}_{i_{n-1} i_n j_{n-1} j_n},
\label{density}
\end{eqnarray}
where $\epsilon^{i_1\cdots i_n}$ is the Levi-Civita symbol. In a two-dimensional manifold the density satisfies $G=\frac{\mathcal{R}_{1212}}{g}=\frac{\mathcal{R}}{2}$. In Ref. \cite{Robson2018}, the authors have shown that the definition of the Euler characteristic integral (\ref{origin1}) is only computed at the Killing horizon of the background spacetime. When the spacetime has more than one Killing horizon, each killing horizon must be considered for the integral (\ref{origin1}).

The Euler characteristic in a two-dimensional spacetime is
\begin{equation}
\chi=\int \sqrt{|g|} d^2x\frac{\mathcal{R}}{4\pi}.
\end{equation}
By using the Wick rotation $t=i\tau$ and setting the new compact time to be the inverse temperature $\beta$, the Euler characteristic $\chi$ becomes \cite{Robson2018}
\begin{equation}
\chi=\int_0^\beta{d\tau}\int_{r_{\text{H}}}\sqrt{|g|}dr \frac{\mathcal{R}}{4\pi}.
\end{equation}
Then the Hawking temperature $T_{H}$ and the Euler characteristic $\chi$ will satisfy the following relation
\begin{equation}
\frac{1}{4\pi T_{H}}\int_{r_{\text{H}}}{\sqrt{|g|}\mathcal{R}dr}=\chi,
\end{equation}
which is the origin of Eq. (\ref{temperature}).

\section{Hawking temperature of the BTZ black hole in general relativity} \label{btz1}

Next, we will use the RVB method to investigate the Hawking temperature of the BTZ black hole in general relativity. The metric of a standard charged rotating BTZ black hole is \cite{Banados1992,Banados1993,Clement1993,Clement1996,Martinez2000}
\begin{equation}
ds^2=-f(r)dt^2+\frac{dr^2}{f(r)}+r^2\left(d\phi-\frac{J}{2r^2}dt\right)^2,
\label{fullmetric}
\end{equation}
where $f(r)$ is the lapse function and reads
\begin{equation}
f(r)=-m+\frac{r^2}{l^2}+\frac{J^2}{4r^2}-2Q^2 \ln \left(\frac{r}{l}\right). \label{fr}
\end{equation}
Here, the parameters $m$, $J$, $l$, and $Q$ correspond to the mass, angular momentum, AdS radius, and charge of the black hole.

\begin{figure}[!htb]
\begin{center}
    \includegraphics[width=0.45\textwidth]{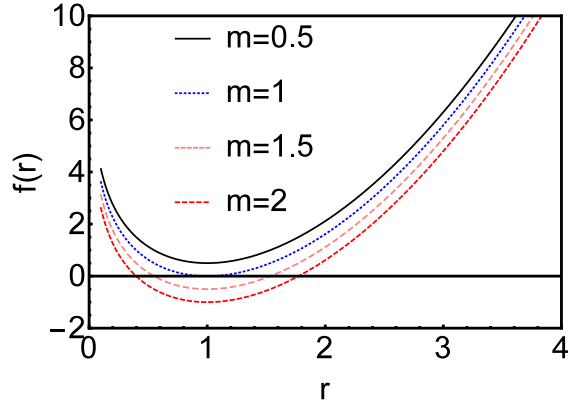}
    \vskip -4mm \caption{Plot of the lapse function $f(r)$ with different mass parameter $m$, where we set the parameters $J=1$, $Q=1$, and $l=1$. The horizons locate at the roots of $f(r)$=0.}
    \label{lapsefunction}
\end{center}
\end{figure}

The value of the lapse function $f(r)$ is dependent on the charge, angular momentum, and mass. When the parameters $l=1$ and $Q=1$, there is no horizon for a sufficiently small mass $m$. This case actually describes a naked BTZ singularity \cite{Hendi:2015wxa}. While when the mass is larger than a certain critical value, the charged BTZ black hole has two horizons, the inner Cauchy horizon $r_{-}$ and the outer event horizon (killing horizon) $r_{+}$, see Fig. \ref{lapsefunction}.

Taking a special hypersurface, the BTZ black hole can be cast into a two-dimensional one with a reduced metric \cite{Achucarro:1993fd} via the Wick rotation ($\tau = i t$):
\begin{equation}
ds^2=f(r)d\tau^2+\frac{dr^2}{f(r)}.
\label{smallmetric}
\end{equation}
The Ricci scalar of the reduced metric (\ref{smallmetric}) reads
\begin{equation}
\mathcal{R}=-\frac{d^2}{dr^2}f(r)=-\frac{2}{l^2}-\frac{3J^2}{2r^4}-\frac{2Q^2}{r^2}.
\end{equation}

\subsection{Charged BTZ black hole ($J=0$, $Q\neq0$)} \label{chargedcasegr}
For the charged BTZ black hole, the lapse function $f(r)$ is given in (\ref{fr}) but with $J=0$ and $Q\neq0$.
When the parameters $l$, $m$, and $Q$ with special fixed values, there are two horizons $r_{-}$ (cauchy horizon) and $r_{+}$ (killing horizon). When the parameters $m$ and $Q$ satisfy some conditions, for examples, $m>1$ and $Q\neq0$, or $m=1$ and $Q>1$, or $m<1$ but large $Q$, there are two horizons $r_{-}$ (cauchy horizon) and $r_{+}$ (killing horizon). So the Euler characteristic of the charged BTZ black hole satisfies $\chi=1$. Accordingly, the Hawking temperature of the black hole can be derived by using Eq.~(\ref{temperature}):
\begin{equation}
T_{H}=\frac{r_{+}}{2l^2\pi}-\frac{Q^2}{2\pi r_{+}},
\label{temperature1}
\end{equation}
which is consistent with the result given in Refs. \cite{Hendi:2015wxa,Tang:2016vmu}.

\subsection{Rotating BTZ black hole ($J\neq 0$, $Q=0$)}
For the rotating BTZ black hole, the lapse function $f(r)$ is
\begin{equation}
f(r)=-m+\frac{r^2}{l^2}+\frac{J^2}{4r^2}.
\end{equation}
When the angular momentum satisfy $J^2 <l^2 m^2$, the rotating BTZ black hole has two horizons
\begin{equation}
 r_{\pm}=\frac{\sqrt{l^2 m\pm\sqrt{l^4 m^2-J^2 l^2}}}{\sqrt{2}},
\end{equation}
where the outer event horizon $r_{+}$ is a killing horizon. Thus the Euler characteristic of the rotating BTZ black hole is $\chi=1$. Substituting the lapse function into Eq. (\ref{temperature}), the Hawking temperature reads
\begin{equation}
T_{H}=\frac{4r_{+}^4-J^2 l^2}{8\pi l^2r_{+}^3},
\end{equation}
which is exactly consistent with the result obtained in Ref. \cite{Setare2007}.

\subsection{Charged rotating BTZ black hole ($J\neq 0$, $Q\neq 0$)}

The charged rotating BTZ black hole also can also have two horizons,
the outer event horizon $r_{+}$ and the Cauchy horzion $r_{-}$. The corresponding Euler characteristic of the charged rotating BTZ black hole is still $\chi=1$. By using the RVB method, the temperature can be calculated as
\begin{equation}
T_{H}=\frac{1}{4\pi}\left(\frac{2r_{+}}{l^2}-\frac{J^2}{2r_{+}^3}-\frac{2Q^2}{r_{+}}\right),
\end{equation}
which is consistent with the result of Ref. \cite{Akbar2011}.

\section{Hawking temperature of the BTZ black hole in massive gravity} \label{btz2}

Next, we will use the RVB method to investigate the Hawking temperature of the BTZ black hole in massive gravity. Firstly, we would like to give a brief review of the BTZ black hole in massive gravity. The three-dimensional massive gravity with a negative cosmological constant and a $U(1)$ gauge field can be described by the following action \cite{Hinterbichler2012,Hendi2016}
\begin{eqnarray}
S &=&-\frac{1}{16\pi}\int d^3x\sqrt{-g}\bigg[\mathcal{R}-2\Lambda +L(F)+\tilde{M}^2\sum_{i=1}^{4}c_iU_i(g,f)\bigg],
\end{eqnarray}
where $L(F)$ is the Lagrangian for the vector gauge field, $\tilde{M}$ is a mass parameter, the parameters $c_i$ are the constants for the massive gravity, $\Lambda=-\frac{1}{l^2}$, and $U_i$ are the eigenvalues of the $d\times d$ matrix $K_\nu^\mu=\sqrt{g^{\mu\alpha}f_{\alpha\nu}}$, which are defined as follows
\begin{eqnarray}
U_1&=&[K],\nonumber\\
U_2&=&[K]^2-[K^2],\nonumber\\
U_3&=&[K]^3-3[K][K^2]+2[K^3],\nonumber\\
U_4&=&[K]^4-6[K^2][K]^2+8[K^3][K]+3[K^2]^2-6[K^4].
\end{eqnarray}
The static solution of the BTZ black hole in massive gravity can also be described by the form of metric (\ref{fullmetric}), while with a different lapse function \cite{Hendi2016}
\begin{equation}
f(r)=-m+\frac{r^2}{l^2}-2Q^2 \ln \left(\frac{r}{l}\right)+\tilde{M}^2cc_1~r.\label{ede}
\end{equation}
The corresponding Ricci scalar for the massive BTZ black hole is
\begin{equation}
\mathcal{R}=-\frac{d^2}{dr^2}f(r)=-\frac{2}{l^2}-\frac{2Q^2}{r^2}.
\label{massivebtzricci}
\end{equation}
Obviously, this Ricci scalar is equal to the case of the charged BTZ black hole in general relativity, which is mainly because that the mass term in (\ref{ede}) has no contribution to the Ricci scalar $\mathcal{R}$. Then by using Eq. (\ref{temperature}), the Hawking temperature of the BTZ black hole in massive gravity can be calculated as
\begin{equation}
T_{H}=\frac{r_{+}}{2l^2\pi}-\frac{Q^2}{2\pi r_{+}},
\label{massivetemperature}
\end{equation}
which is the same as the case of the charged BTZ black hole given in Section \ref{chargedcasegr}. However, the Hawking  temperature of the BTZ black hole in massive gravity is \cite{Chougule:2018cny}
\begin{equation}
T_{H}=\frac{r_{+}}{2l^2\pi}-\frac{Q^2}{2\pi r_{+}}-\frac{\tilde{M}^2 cc_1}{4\pi}.
\label{massiverighttemperature}
\end{equation}
Comparing (\ref{massivetemperature}) and (\ref{massiverighttemperature}), one will find that the temperature obtained from the RVB method is quite different from the Hawking temperature for a massive BTZ black hole. Therefore, it seems that the RVB method fails for the massive BTZ black hole. However, we should note that there exists an
integral constant when one performs the integral (\ref{temperature}). Keeping this in mind, we can add an integral constant $c=-\frac{\tilde{M}^2 cc_1}{4\pi}$ into Eq. (\ref{massivetemperature}), then we will get the correct Hawking temperature (\ref{massiverighttemperature}) for the massive BTZ black hole. In summary, the Hawking temperature can be exactly reproduced following the RVB method by including the integral constant.

On the other hand, in order to avoid the integral constant problem, we can perform the following simple calculation. Plugging Eq. (\ref{massivebtzricci}) into Eq. (\ref{temperature}), we can get
\begin{eqnarray}
 T_{\text{H}}&=&\frac{1}{4\pi \chi}\int_{r_{+}}{\frac{d^2}{dr^2}f(r)dr}\nonumber\\
 &=&\frac{1}{4\pi \chi}\int_{r_{+}}{df'(r)}\nonumber\\
 &=&\frac{f'(r_+)}{4\pi \chi},
\end{eqnarray}
for a two-dimensional hypersurface given in the $\tau$-$r$ plane. By taking $\chi=1$, we will the result of the temperature that follows the standard `Euclidean trick'. Therefore, the RVB method can exactly reproduce the Hawking temperature for the black hole of the reduced metric (\ref{smallmetric}). So it is an universal result for the RVB method.

\section{Summary and Conclusions} \label{Conclusion}

In this paper, firstly we gave a brief introduction and review of the RVB method. The Hawking temperature of the charged and rotating BTZ black hole in general relativity can be exactly obtained by using the RVB method with the topological formula (\ref{temperature}). Here, we need to note that the Euler characteristic is defined in the even-dimensional manifold, and the three-dimensional BTZ black hole corresponds to the odd-dimensional manifold. This implies that even in the three-dimensional spacetime, we can still obtain the temperature of the black hole with a reduced two-dimensional hypersurface via the RVB method.

We also investigated the Hawking temperature of the BTZ black hole in massive gravity. We found that if we directly use the Ricci scalar (\ref{massivebtzricci}) to calculate the Hawking temperature, the corresponding result will lose the information given by the mass term $\tilde{M}^2cc_1~r$. It seems that the RVB method could not give the proper temperature. However, if we include the integral constant term, the exact Hawking temperature will be obtained. Moreover, directly plugging the Ricci scalar (\ref{massivebtzricci}) into the integral (\ref{temperature}), the Hawking temperature can be exactly obtained, which implies that the RVB method is an universal approach to obtain the Hawking temperature for a black hole.

\section*{Acknowledgments}

This work was supported in part by the National Natural Science Foundation of China (Grants No. 11875151, No. 11675064, and No. 11522541), the Fundamental Research Funds for the Central Universities (Grants No. lzujbky-2018-k11 and No. lzujbky-2017-it68). Y.P. Zhang was supported by the scholarship granted by the Chinese Scholarship Council (CSC).

\end{document}